\documentclass[aps,twocolumn,pre, superscriptaddress]{revtex4-2}

\usepackage[english]{babel}
\usepackage{amsmath, amssymb, mathtools, physics}
\usepackage{braket}
\usepackage{ifthen}
\usepackage{graphicx,float}
\usepackage{multirow}
\usepackage[RGB,dvipsnames]{xcolor}
\definecolor{dodgerblue}{rgb}{0.12, 0.56, 1.0}
\definecolor{darkcyan32144140}{RGB}{32,144,140}
\definecolor{darkslateblue5294141}{RGB}{52,94,141}      % 2
\definecolor{darkslateblue6467135}{RGB}{64,67,135}
\definecolor{darkslateblue7235116}{RGB}{72,35,116}      % 5
\definecolor{darkslategray38}{RGB}{38,38,38}
\definecolor{greenyellow18922238}{RGB}{189,222,38}      % 4
\definecolor{indigo68184}{RGB}{68,1,84}                 % 1
\definecolor{lavender234234242}{RGB}{234,234,242}
\definecolor{mediumseagreen34167132}{RGB}{34,167,132}   % 3
\definecolor{mediumseagreen68190112}{RGB}{68,190,112}
\definecolor{teal41120142}{RGB}{41,120,142}
\definecolor{yellowgreen12120981}{RGB}{121,209,81}
\colorlet{bluedarkslateblue5294141}{blue!50!darkslateblue5294141}

\newcommand{\eprBinned}{\sigma_{\text{WTD}}^{\Delta t}}
\newcommand{\eprBinnedhat}{\hat{\sigma}_{\text{WTD}}^{\Delta t}}
\newcommand{\eprM}{\sigma_\text{M}}

\newcommand{\eprWTD}{\sigma_\text{WTD}}
\newcommand{\eprWTDhat}{\hat{\sigma}_\text{WTD}}
\newcommand{\eprEMC}{\sigma_\text{EMC}}
\newcommand{\eprEMChat}{\hat{\sigma}_\text{EMC}}

\newcommand{\mean}[1]{\left\langle #1\right\rangle}
\newcommand{\wtdpsi}[3]{\psi_{#1\to #2}(#3)}
\newcommand{\empwtdpsi}[3]{\hat{\psi}_{#1\to #2}(#3)}
\newcommand{\binnedwtdpsi}[3]{\Psi_{#1\to #2}(#3)}
\newcommand{\empbinnedwtdpsi}[3]{\hat{\Psi}_{#1\to #2}(#3)}
\newcommand{\uboundbinned}{I^{\ast,\Delta t}}

\usepackage{hyperref}
\usepackage[nameinlink]{cleveref}
\usepackage{enumitem}
\usepackage{tikz}
\usetikzlibrary{external,cd}
\usepackage{pgfplots}
\renewcommand{\vec}{\boldsymbol}
\usetikzlibrary{automata, arrows.meta, positioning, calc,shapes}

\begin{document}
%==============================[ Definitionen der Titelseite. ]==============================%
%
\title{A pedestrian's approach to large deviations in semi-Markov processes with an application to entropy production}
\author{Alexander M. Maier}
\thanks{These authors contributed equally}
\author{Jonas H. Fritz}
\thanks{These authors contributed equally}
\author{Udo Seifert}

\affiliation{%
 II. Institut für Theoretische Physik, Universität Stuttgart, 70550 Stuttgart, Germany
}%
\date{\today}
% \date{\today} % TODO
% \keywords{}

\begin{abstract}
Semi-Markov processes play an important role in the effective description of partially accessible systems in stochastic thermodynamics. They occur, for instance, in coarse-graining procedures such as state lumping and when analyzing waiting times between few visible Markovian events. The finite-time measurement of any coarse-grained observable in a stochastic system depends on the specific realization of the underlying trajectory. Moreover, the fluctuations of such observables are encoded in their rate function that follows from the rate function of the empirical measure and the empirical flow in the respective process. Derivations of the rate function of empirical measure and empirical flow in semi-Markov processes with direction-time independence (DTI) exist in the mathematical literature, but have not received much attention in the stochastic thermodynamics community. We present an accessible derivation of the rate function of the tuple frequency in discrete-time Markov chains and extend this to the rate function of the empirical semi-Markov kernel in semi-Markov processes without DTI. From this, we derive an upper bound on the rate function of the empirical entropy production rate, which leads to a lower bound on the variance of the mean entropy production rate measured along a finite-time trajectory. We illustrate these analytical bounds with simulated data.
\end{abstract}
%==============================[ Definitionen der Titelseite. Ende. ]==============================%
\maketitle

%================================================================================================================================[ SECTION ]===========%
\section{Introduction}
Large deviation theory \cite{vara84,holl00,demb09,touc09,touc18} is a mathematical framework that has proven to be invaluable in stochastic thermodynamics \cite{seki10,jarz11,vdb15,peli21,shir23,seif25,maes08,ging16,ging17}. In this theory, the rate function plays a crucial role. It quantifies how fast deviations from the mean decay in the asymptotic limit. In the mathematical literature, the rate function has been derived for independent and identically distributed processes \cite{sano57,csis98}, discrete-time Markov chains \cite{nata85,holl00,vidy09}, and continuous-time Markov chains \cite{bert15}. For semi-Markov networks with direction-time independence (DTI), in which waiting-time distributions factor into a transition probability and a sojourn-time distribution, a fluctuation theorem for the current \cite{andr08} and the rate function for waiting-time distributions that are allowed to have heavy tails \cite{mari16,jia22} have been derived. Even though semi-Markov processes with DTI have emerged as a description of certain state-lumped Markov networks \cite{wang07a,mart19,teza20,skin21a,erte22}, the rate function of waiting-time distributions in such processes has received little attention in the stochastic thermodynamics community so far \cite{sugh18}. This might be because extant derivations are rather formal and technical. Additionally, large deviations of semi-Markov processes without DTI have not been studied rigorously. The first goal of this work is to close these gaps by providing an accessible derivation of the rate function of semi-Markov processes without DTI.

An important realization of such processes is the waiting-time-based description of a partially accessible Markov network. These semi-Markov processes allow inferring lower bounds on the entropy production of underlying Markov networks with only a few visible transitions \cite{vdm22,haru22,piet24,maie24,erte24,haru24,igos25a} and of more general setups that go beyond the Markov assumption \cite{vdm22b,degu24,degu24a}. Recently, challenges in applying these estimators to experimental data have been studied \cite{gode22,voll24,song24,frit25}. Beyond limited access to only a few degrees of freedom \cite{blom24}, rare events might not be measured at all in a finite measurement time \cite{baie24} and detectable transitions might be misidentified \cite{vdm25}.

Experiments have limited temporal resolution and finite measurement statistics. Recent works have studied how this affects the entropy production rate by calculating its large-deviation rate function directly for specific Langevin dynamics \cite{chen16,budh21}. The second goal of this work is to study the large deviations of the entropy (estimator) introduced in \cite{haru22,vdm22}, which relies on waiting-time distributions between detectable transitions of a partially observable Markov network. The rate function of this entropy estimator is formally defined as the contraction of the rate function of waiting-time distributions of the corresponding semi-Markov network.

This paper is structured as follows. We introduce semi-Markov dynamics and present a pedestrian's derivation of the rate functions of tuple frequencies in discrete-time Markov chains and of waiting-time distributions in semi-Markov networks without DTI in \Cref{sec:LDP_SM}. A rigorous proof and an explanation of how this rate function can be evaluated are provided in more detail in \Cref{sec:proof}. In \Cref{sec:LDP_EPR}, we derive the rate functions of different entropy productions: first in embedded Markov chains (EMC) and then in semi-Markov processes. We demonstrate our results with data from a Gillespie simulation. Finally, we conclude in \Cref{sec:conclusion}.

%================================================================================================================================[ SECTION ]===========%
\section{Large deviations of semi-Markov processes}
\label{sec:LDP_SM}
%================================================================================================================================[ SUBSECTION ]--------%
\subsection{Semi-Markov process and embedded Markov chain}
\begin{figure*}
\centering
% \tikzsetnextfilename{ld_graph}
% \input{tikz/ld_graph.tikz}
\includegraphics[scale=1]{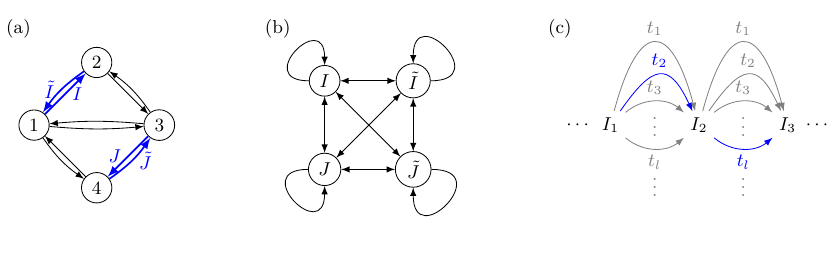}
\caption{Semi-Markov networks. (a) Graph of a partially accessible 4-state Markov network. The four transitions $I,\tilde{I},J$ and $\tilde{J}$ (blue) are observable, whereas states and other transitions are invisible, leading to: (b) Graph of the corresponding 4-state semi-Markov network. (c) Sequence of semi-Markov states along a section of a sample trajectory and the concept of channels. By treating waiting times between consecutive states as finely discretized, each waiting time $t_l$ corresponds to one of infinitely many channels between two successive states. The probability of passing through a channel between two consecutive states is thereby given by the corresponding discrete waiting-time distribution. In this specific example, the trajectory passes through blue channels.
}
\label{fig:ldgraphfig}
\end{figure*}

We consider a semi-Markov process with states $I,J,\dots$ that is defined by the set of its waiting-time distributions $\left\{\psi_{I\to J}(t)\right\}$, also called semi-Markov kernel. A waiting-time distribution \begin{equation}
    \psi_{I\to J}(t) \equiv p[J;t_{J} - t_{I}=t|I]
    \label{eq:wtd_def}
\end{equation}
is the probability of observing the event $J$ after time $t$, given that event $I$ has occurred at time $t_I$.
Since reaching a state $I$ is a renewal event, which means that the probability \eqref{eq:wtd_def} is independent of the trajectory that precedes $I$, definition \eqref{eq:wtd_def} depends on the waiting time $t$ but is independent of the specific time $t_I$ at which $I$ is observed. The same applies to the dynamics under time reversal, which we denote throughout by a tilde. Moreover, in contrast to the usual case, the semi-Markov states may be odd under time reversal such that $\tilde{\tilde{I}}=I$ with $\tilde{I}\neq I$. Such states emerge, for example, when we can only observe few pairs of transitions of a Markov network as it is the case in \Cref{fig:ldgraphfig}\,(a). The associated semi-Markov network of this example with states that arise from transitions is displayed in \Cref{fig:ldgraphfig}\,(b).

If the waiting-time distribution \eqref{eq:wtd_def} factorizes into the transition probability $p_{IJ}$ and a transition-independent waiting time $\psi_I(t)$,
\begin{equation}
    \psi_{I\to J}(t) = p_{IJ} \psi_I(t)\,,
    \label{eq:DTI_def}
\end{equation}
the process is part of the special subclass of semi-Markov processes that have direction-time independence (DTI). Whereas DTI is a common assumption on semi-Markov processes in the literature \cite{wang07a,andr08,maes09,mari16,vanv20a,jia22}, the processes we consider throughout this paper typically do not have DTI.

By disregarding the time between consecutive events and by only considering the sequence of events, we end up with a simplified description of such a process, the so-called embedded Markov chain (EMC). The EMC is fully characterized by its transition probabilities
\begin{equation}
    p_{IJ} = \int_{0}^{\infty}\dd{t} \psi_{I\to J}(t)\,,
    \label{eq:trans_prob_def}
\end{equation}
for jumps from state $I$ to $J$, which define the transition matrix $(\vec{P})_{JI}=p_{IJ}$. Through the eigenvalue problem
\begin{equation}
    \vec{\pi} = \vec{P}\vec{\pi}\,,
    \label{eq:pi_def}
\end{equation}
it defines the steady-state probability of the EMC with components $\pi_I$. The probability of observing the tuple $IJ$ in the steady state of the EMC is then given by
\begin{equation}
    \pi_{IJ} = \pi_Ip_{IJ}\,.
    \label{eq:pi_p}
\end{equation}
which implies
\begin{equation}
    \sum_J \pi_{IJ}=\pi_I.
    \label{eq:pi_I_from_pi_IJ}
\end{equation}
Moreover, we denote the average frequency of registering an event $I$ in the original semi-Markov process by 
\begin{equation}
    \nu_I=\pi_I/\braket{t}\,,
\end{equation}
where $\braket{t}$ is the average waiting time between two events.

\subsection{Basic concepts in large deviations theory}

The fluctuations of a time-intensive quantity, like the entropy production rate, can be studied using the framework of large deviations. In this framework, the probability distribution of such a general, time-intensive empirical quantity $\hat o$ scales as 
\begin{equation}
    p(\hat o) \asymp \exp\left[-T I(\hat o)\right]\,
    \label{eq:I_def}
\end{equation}
for large times $T$, where $I(\hat o)$ is the so-called rate function. As for $\hat{o}$, we will use a hat to denote empirically measured quantities throughout this work. The minimum of this rate function corresponds to the mean value $\mean{o}$, with $I(\braket{o})=0$. Expanding the rate function around its minimum yields the variance of $\hat o$,
\begin{equation}
    \text{var}\left(\hat o\right) = 1/TI^{\prime\prime}(\braket{o})\,,
    \label{eq:var_from_I}
\end{equation}
where the primes represent the second derivative.

The quantity $\hat o$ we are interested in may depend on some underlying quantities $\vec{\hat\zeta}$, such that $\hat o=o(\vec{\hat\zeta})$. If we know the rate function of $\vec{\hat\zeta}$, we obtain the rate function of $\hat o$ as the contraction \cite{vara84,holl00,demb09,touc09,touc18}
\begin{align}
I(\hat{o}) = \min_{\vec{\hat\zeta}} \left\lbrace I\left(\vec{\hat\zeta}\right) \middle\vert o\!\left(\vec{\hat\zeta}\right) = \hat{o}\right\rbrace,
\label{eq:Icontraction_o}
\end{align}
where the minimization is taken over all $\vec{\hat \zeta}$ that lead to the value $\hat o$. 

%======================================================================================================================[ FIGURE ]
%==================================================================================================================[ FIGURE END ]

\subsection{A pedestrian's path to the rate function}
The rate function of fluxes in discrete-time Markov chains and semi-Markov processes has been derived using different methods in the mathematical literature \cite{nata85,vidy09,mari16,jia22}. Here, we present a concise derivation that is immediately accessible. We start by considering an EMC with transition probabilities $\left\{p_{IJ}\right\}$ and trajectories of length $N$ starting at $I_0$. For each trajectory, we count the number of occurrences of each tuple $IJ$, which we denote by $\hat{n}_{IJ}$. The probability of observing a set $\left\{\hat{n}_{IJ}\right\}$ is given by \begin{equation}
    p\left(\left\{\hat{n}_{IJ}\right\}\vert I_0\right) = \mathcal{N}(\left\{\hat{n}_{IJ}\right\})\prod_{IJ}p_{IJ}^{\hat{n}_{IJ}}\,,
    \label{eq:ped_normal}
\end{equation}
where $\mathcal{N}(\left\{\hat{n}_{IJ}\right\})$ is the number of different trajectories that lead to the observation of the set $\left\{\hat{n}_{IJ}\right\}$. The typical observations in this dynamics for large $N$, $\left\{n_{IJ}\right\}=\left\{\pi_{IJ}N\right\}$ and $\left\{n_{I}\right\}=\left\{\pi_{I}N\right\}$, are related by \eqref{eq:pi_p} implying
\begin{equation}
    p_{IJ}=\pi_{IJ}/\pi_I=n_{IJ}/n_I\,.
    \label{eq:p_IJ_from_n_IJ}
\end{equation}

We introduce an auxiliary dynamics with the same graph but different transition probabilities $\left\{\hat{p}_{IJ}\right\}$, in which $\left\{\hat{n}_{IJ}\right\}$ is the typical observation. We then have
\begin{equation}
    \hat{p}\left(\left\{\hat{n}_{IJ}\right\}\vert I_0\right) = \mathcal{N}(\left\{\hat{n}_{IJ}\right\})\prod_{IJ}\left(\hat{p}_{IJ}\right)^{\hat{n}_{IJ}}\,.
    \label{eq:ped_rescaled}
\end{equation}
Additionally, the analog of \eqref{eq:p_IJ_from_n_IJ} is
\begin{equation}
    \hat{p}_{IJ}=\hat{\pi}_{IJ}/\hat{\pi}_I=\hat{n}_{IJ}/\hat{n}_I\,.
    \label{eq:p_dagger_IJ_from_n_IJ}
\end{equation}
Moreover, solving \Cref{eq:ped_rescaled} for $\mathcal{N}(\left\{\hat{n}_{IJ}\right\})$ and inserting this into \Cref{eq:ped_normal} yields
\begin{equation}
    p\left(\left\{\hat{n}_{IJ}\right\}\vert I_0\right) = \prod_{IJ}\left(\frac{p_{IJ}}{\hat{p}_{IJ}}\right)^{\hat{n}_{IJ}}\hat{p}\left(\left\{\hat{n}_{IJ}\right\}\vert I_0\right).
    \label{eq:ped_inserted}
\end{equation}

Close to its mean, we can approximate the distribution $\hat{p}\left(\left\{\hat{n}_{IJ}\right\}\right)$ as Gaussian
\begin{equation}
    \hat{p}\left(\left\{\hat{n}_{IJ}+\delta_{IJ}\right\}\vert I_0\right) \sim 1/N^{d}\exp\left[-\frac{1}{2N}\left(\vec{\delta^T\Sigma^{-1}\delta}\right)\right]\,
    \label{eq:ped_gaussian}
\end{equation}
where the covariance matrix $\vec{\Sigma}^{-1}$ of the time-extensive $\vec{\delta}$ scales as $1/N$. Thus, the normalization of this distribution is proportional to $1/N^{d}$, where $d$ is the dimension of $\vec{\delta}$, i.e., the number of different tuples $IJ$. In the limit $\vec{\delta}\to\vec{0}$, we thus find $\hat{p}\left(\left\{\hat{n}_{IJ}\right\}\vert I_0\right)\sim 1/N^d$, which is neither exponentially small nor large. Hence, from \eqref{eq:ped_inserted} we obtain
\begin{equation}
    p\left(\left\{\hat{n}_{IJ}\right\}\right)\asymp \exp\left[-\left(\sum_{IJ}\hat{n}_{IJ}\ln\frac{\hat{n}_{IJ}}{n_{IJ}}-\sum_{I}\hat{n}_I\ln\frac{\hat{n}_I}{n_I}\right)\right]
    \label{eq:p_asympt_nIJ}
\end{equation}
after dropping the now irrelevant $I_0$.
The exponent in \eqref{eq:p_asympt_nIJ} scales with $N$. Hence, dividing the exponent by $N$ yields a rate function similar to the case in \eqref{eq:I_def}, where $I(\hat{o})$ is independent of $T$. Using \Cref{eq:p_IJ_from_n_IJ,eq:p_dagger_IJ_from_n_IJ} again, we arrive at the rate function
\begin{equation}
   I\left(\left\{\hat{p}_{IJ}\right\}\right)= \sum_{IJ}\hat{\pi}_I\hat{p}_{IJ}\ln\frac{\hat{p}_{IJ}}{p_{IJ}}\,,
   \label{eq:Iemc}
\end{equation}
expressed as a function of the empirical conditional probabilities $\left\{\hat{p}_{IJ}\right\}$. Equivalently,
\begin{equation}
    I\left(\left\{\hat{\pi}_{IJ}\right\}\right) = \sum_{IJ}\hat{\pi}_{IJ}\ln\frac{\hat{\pi}_{IJ}}{\pi_{IJ}}-\sum_{I}\hat{\pi}_I\ln\frac{\hat{\pi}_I}{\pi_I}
    \label{eq:I_tuple}
\end{equation}
is the rate function of the unconditioned probabilities $\left\{\hat{\pi}_{IJ}\right\}$.
This result is well established in the mathematical literature \cite{nata85,holl00,demb09}. For a pedagogical derivation from a mathematical background see \cite{vidy09}.

The rate function of the semi-Markov kernel can be obtained by a generalization of this derivation. We introduce channels labeled by $l$ that correspond to a discretized waiting time $t_l$ between two consecutive states $I$ and $J$. The probability for a transition from $I$ to $J$ along such a channel with $t_l$ given the observation of $I$ at time $0$ results from binning the corresponding waiting-time distribution as
\begin{equation}
    p_{IJ,l}\equiv\binnedwtdpsi{I}{J}{t_l} \equiv\frac{1}{\Delta t}\int_{t_l-\Delta t/2}^{t_l+\Delta t/2}\dd{t} \wtdpsi{I}{J}{t}.
    \label{eq:psi_dis_def}
\end{equation}
\Cref{fig:ldgraphfig}\,(c) displays a section of an illustrative sample trajectory. Instead of counting $\hat{n}_{IJ}$ as above, we now count the number of transitions $I\to J$ along a channel $l$, $\hat{n}_{IJ,l}$, which greatly increases the number of observables.
Performing steps that are analogous to the ones for the EMC, just treating $l$ as a further index, we obtain 
\begin{align}
I\left[\left\{\empbinnedwtdpsi{I}{J}{t_l}\right\}\right] = \sum_{IJ,l} \hat{\nu}_I \empbinnedwtdpsi{I}{J}{t_l}\ln\frac{\empbinnedwtdpsi{I}{J}{t_l}}{\binnedwtdpsi{I}{J}{t_l}}\,.
\label{eq:I_binnedpsi}
\end{align}
By finally taking the limit $\Delta t\rightarrow 0$, the sum turns into an integral, such that 
\begin{equation}
    I\left[\left\{\empwtdpsi{I}{J}{t}\right\}\right] = \sum_{IJ}\int\dd t\, \hat{\nu}_I \empwtdpsi{I}{J}{t}\ln\frac{\empwtdpsi{I}{J}{t}}{\wtdpsi{I}{J}{t}}\,.
    \label{eq:I_wtdneu}
\end{equation}

The rate function \eqref{eq:I_wtdneu} superficially looks like a Kullback-Leibler divergence, which is the form of extant results in mathematical literature, cf. \cite{mari16,sugh18,jia22}. To rigorously prove this rate function, however, we still need to show that our initial assumptions regarding the infinite number of transition channels and the effective binning are admissible and that the limits are well behaved. Thus, we turn to the rigorous derivations in \cite{mari16,jia22} for semi-Markov processes with DTI and use these results to provide a rigorous derivation for the rate function \eqref{eq:I_wtdneu} in \Cref{sec:proof}. We thus provide proof that our easy-to-follow pedestrian's approach, which does not need the assumption of DTI, leads to the correct rate functions \eqref{eq:I_tuple} and \eqref{eq:I_wtdneu}.

%================================================================================================================================[ SECTION ]===========%
\section{Large deviations of entropy production rate}
\label{sec:LDP_EPR}
%================================================================================================================================[ SUBSECTION ]--------%
\subsection{Entropy production from waiting-time distributions}
An important type of a semi-Markov process emerges as a coarse-grained description of a partially accessible Markov network with states $i,j,\dots$, in which only certain transitions $i\to j$ can be observed, see e.g. \cite{vdm22,haru22}. In this case, the states of the resulting semi-Markov network are transitions in the underlying Markov network, e.g. $I=i \to j$. They are thus odd under time reversal like $I \neq \tilde{I}$ in \Cref{fig:ldgraphfig} (a) and (b).

For the underlying Markov process with steady state distribution $\vec{\pi}$ and transition rates $k_{ij}$, the mean entropy production rate is given by \cite{vdb15,peli21,shir23,seif25}
\begin{equation}
    \eprM = \sum_{ij}\pi_ik_{ij}\ln\frac{\pi_ik_{ij}}{\pi_j k_{ji}}\,.
    \label{eq:sig_m_def}
\end{equation}
The entropy production rate of the semi-Markov process is \cite{vdm22,haru22}
\begin{equation}
    \eprWTD\equiv \sum_{IJ}\int_0^{\infty}dt\,\nu_I\psi_{I\to J}(t)\ln\frac{\psi_{I\to J}(t)}{\psi_{\widetilde{J}\to \widetilde{I}}(t)}\,,
    \label{eq:sig_wtd_def}
\end{equation}
taking into account the parity of the states. It follows from a fluctuation theorem for path weights by using the log-sum inequality and is thus a thermodynamically consistent estimator for the entropy production in the underlying Markov network, in the sense that $\eprWTD\leq \eprM$. Applying the log-sum inequality once again to \eqref{eq:sig_wtd_def} leads to the entropy production rate of the EMC,
\begin{align}
    \eprEMC \equiv \sum_{IJ}\nu_{I}p_{I\to J}\ln\left(\frac{p_{I\to J}}{p_{\tilde{J}\to\tilde{I}}}\right)\leq \eprWTD.
    \label{eq:sig_emc_def}
\end{align}
%================================================================================================================================[ SUBSECTION ]--------%
\subsection{Embedded Markov chain}
\label{sec:LDP_EPR_EMC}
%======================================================================================================================[ FIGURE ]
\begin{figure*}
\centering
\includegraphics[scale=1]{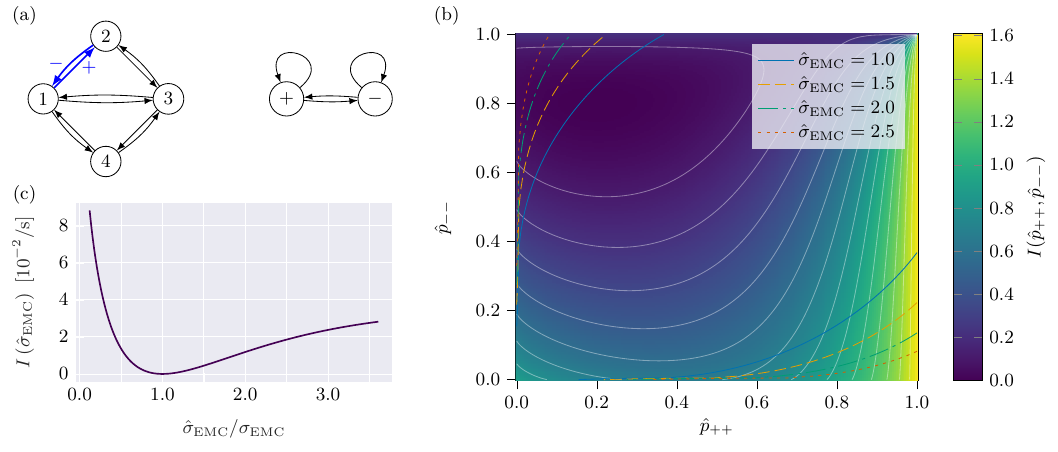}
% \tikzsetnextfilename{emc_figure}
% \input{tikz/emc_figure.tikz}
\caption{
Large-deviation rate functions of a 2-state EMC. (a) 4-state Markov network, in which only transitions $+$ and $-$ are observable, and the graph of the associated semi-Markov network and its EMC. (b) Heat map with contour lines of the rate function \eqref{eq:Iemc} depending on the two empirical transition probabilities $\hat{p}_{++}$ and $\hat{p}_{--}$ of the EMC shown in (a). The four pairs of lines with the same distinct line style each indicate the combinations of $\hat{p}_{++}$ and $\hat{p}_{--}$ that yield the same value of $\eprEMChat$. (c) Rate function of the entropy production $\eprEMChat$ resulting as the contraction of $I(\hat{p}_{++},\hat{p}_{--})$ displayed in (b). Finding the value of the rate function for the four values of $\eprEMChat$ in (b) is equivalent to determining the minimum of $I(\hat{p}_{++},\hat{p}_{--})$ along the corresponding lines in the heat map. For this figure, $p_{++}=0.2$ and $p_{--}=0.8$ fix the transition probabilities of the EMC.
}
\label{fig:Iemc}
\end{figure*}
%==================================================================================================================[ FIGURE END ]
We first discuss the rate function of the entropy production rate $\eprEMC$ on the level of the EMC.
Hence, we aim to determine the contraction \eqref{eq:Icontraction_o} of the rate function \eqref{eq:Iemc} using $o=\eprEMC$.
The minimization in \eqref{eq:Icontraction_o} must thus be performed subject to the condition
\begin{align}
\eprEMC\left(\left\lbrace \hat{p}_{IJ}\right\rbrace\right) = \hat{\sigma}_\text{EMC}\,. \label{eq:EMCcond}
\end{align}

For an illustration, we use the simplest possible setup, a system of two states $+$ and $-$ that are each others time reverse in an underlying Markov network. In this system, the two transition probabilities $p_{++}$ and $p_{--}$ fix the other two, $p_{+-} = 1- p_{++}$ and $p_{-+} = 1-p_{--}$, and the steady state is $\pi_+ = (p_{--} -1)/(p_{--}+p_{++}-2) = 1-\pi_-$. Moreover, using a Lagrange multiplier for the minimization with $0<p_{\pm\pm}<1$ leads to transcendental equations
\begin{align}
    (\partial_{\hat{p}_{++}}I)\partial_{\hat{p}_{--}}\eprEMC\left(\left\lbrace \hat{p}_{IJ}\right\rbrace\right) &= (\partial_{\hat{p}_{--}}I)\partial_{\hat{p}_{++}}\eprEMC\left(\left\lbrace \hat{p}_{IJ}\right\rbrace\right) \label{eq:EMC_transeqii}
\end{align}
and condition \eqref{eq:EMCcond}. Solving this set of equations analytically is impossible in general. In Figure \ref{fig:Iemc}, we illustrate the numerical minimization and the resulting rate function in the case of a 4-state Markov network.

%================================================================================================================================[ SUBSECTION ]--------%
\subsection{Semi-Markov process and discretized waiting-time distributions}
\begin{figure*}
    \centering
    % \tikzsetnextfilename{sM_figure}
    % \input{tikz/sM_figure.tikz}
    \includegraphics[scale=1]{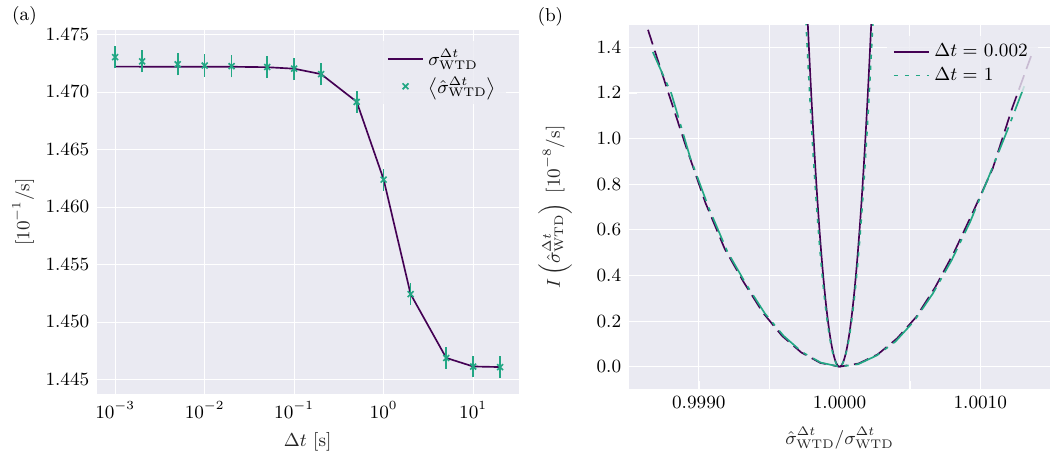}
    \caption{Empirically measured entropy estimator and its large deviations from over $160\thinspace000$ trajectories of the semi-Markov network introduced in \Cref{fig:Iemc}. (a) The dependence of $\eprBinned$ on $\Delta t$. The value of $\eprBinned$ for perfect statistics (purple) is approximately constant for binnings much finer than the characteristic time scale of the system and decreases for coarser time resolutions, which decreases its quality as an estimator of $\eprM$. Crosses and bars indicate the mean and the 2-standard-deviation width of the empirical distribution at each $\Delta t$, respectively. Hence, $\eprBinnedhat$ may under- or overestimate the semi-Markov entropy production for any finite set of samples. For increasing $\Delta t$, the mean of $\eprBinnedhat$ converges faster to its analytical value while the width of the distributions remains relatively constant. (b) The empirical rate function (dashed and dash-dotted lines) and their upper bounds (solid and dotted lines) for two different $\Delta t$. Due to the length of the simulated trajectories, only a small region around the mean can be sampled. Methods: The rates of the underlying Markov network are $k_{12}=k_{23}=k_{31}=2$, $k_{14}=k_{43}=3$ and for all other transitions $k_{ij}=1$. The trajectories of the semi-Markov process with states $+$ and $-$ have been generated in analogy to Gillespie's algorithm \cite{gill77} by randomly drawing two random numbers in each step, one determines the next state and the other determines the waiting time until this state is visited based on the corresponding cumulative waiting-time distribution. The generated trajectories have length $T=6\cdot 10^8$ to ensure that large deviation theory is applicable. We record the binned waiting-time distributions at a fine resolution of $\Delta t=10^{-3}$ with a cutoff time $t_{\text{max}}=20$.} %169314
    \label{fig:wtd}
\end{figure*}
As discussed in \Cref{sec:LDP_SM}, we evaluate the rate function \eqref{eq:I_wtdneu} using binned waiting-time distributions \eqref{eq:psi_dis_def}. Evaluating \Cref{eq:sig_wtd_def} with these discretized waiting-time distributions, which corresponds to applying the log-sum inequality to \eqref{eq:sig_wtd_def}, we arrive at the entropy production rate
\begin{align}
    \eprBinned \equiv  \sum_{I,J,l}\nu_{I} \binnedwtdpsi{I}{J}{t_l} \ln\frac{\binnedwtdpsi{I}{J}{t_l}}{\binnedwtdpsi{\widetilde{J}}{\widetilde{I}}{t_l}} \leq \eprWTD\,.
    \label{eq:binned_sig_wtd_def}
\end{align}

We show in \Cref{fig:wtd}\,(a) for the Markov network from \Cref{fig:Iemc} that this binning leaves $\eprBinned$ approximately constant as long as it is much finer than the characteristic time scale of the system. Moreover, this figure illustrates the probabilistic nature of $\eprBinnedhat$ through the mean and the 2-standard-deviation width of its empirical distribution for each $\Delta t$, which we get by simulating a large number of trajectories of the semi-Markov process based on the Markov network shown in \Cref{fig:Iemc}.
As we might expect for increasing $\Delta t$, the empirical mean converges faster to its analytical value, as do the waiting-time distributions at coarser resolutions.

Next, we seek to gain analytical insight into the fluctuations of $\eprWTDhat$ and $\eprBinnedhat$  by turning to their rate functions, for the latter of which binning leads to additional subtleties. This rate function is formally given as the contraction \eqref{eq:Icontraction_o} of the rate function \eqref{eq:I_binnedpsi}, using $\hat o=\eprBinnedhat$ and minimizing over $\vec{\hat\zeta} = \left\{\empbinnedwtdpsi{I}{J}{t_l}\right\}$, which, however, is infeasible in practice.
Nevertheless, an upper bound on the rate function $I(\eprBinnedhat)$ results from any set of waiting-time distributions $\left\{\hat{\Psi}_{I\to J}^{\ast}(t_l)\right\}$ that satisfies the condition 
\begin{equation}
    \eprBinned\left(\left\{\hat{\Psi}_{I\to J}^{\ast}(t_l)\right\}\right)=\eprBinnedhat\,.
    \label{eq:sM_epr_rescale_cond}
\end{equation}
We thus need to find a set of waiting-time distributions that fulfills this constraint. Inspired by Refs. \cite{ging16,garr17}, we choose a simple rescaling of the time scale in the waiting-time distribution, 
\begin{equation}
    \hat{\Psi}_{I\to J}^{\ast}(t_l) \equiv s\binnedwtdpsi{I}{J}{st_l}=\frac{1}{\Delta t}\int_{t_l-\Delta t/2}^{t_l+\Delta t/2}\dd{t} s\wtdpsi{I}{J}{st}\,.
    \label{eq:psi_rescale}
\end{equation}
For this rescaling, we only scale the underlying continuous distribution, keeping the bin width $\Delta t$ and the integration bounds, i.e., the discretization, unchanged. This ensures that we can compare the rescaled histograms with the original ones. Additionally, this scaling leaves the transition probabilities and thus the steady-state distribution $\vec{\pi}$ unchanged, which we need due to the intractable dependence of $\vec{\pi}$ on $\left\{\psi_{IJ}(t)\right\}$. 
Choosing
\begin{equation}
    s(\eprBinnedhat)=\eprBinnedhat/\eprBinned
\end{equation}
leads to the desired entropy production \eqref{eq:sM_epr_rescale_cond} and we get $\nu_I^*=s\nu_I$. Inserting this and the rescaled distributions \eqref{eq:psi_rescale} into the rate function \eqref{eq:I_binnedpsi} yields the following bound on the rate function 
\begin{align}
    I^{}(\eprBinnedhat)\leq& \sum_{I,J,l}\,s^2\nu_I \Delta t\binnedwtdpsi{I}{J}{st_l}\ln\left[\frac{s\binnedwtdpsi{I}{J}{st_l}}{\binnedwtdpsi{I}{J}{t_l}}\right]\\
    &\equiv \uboundbinned\!\left[s(\eprBinnedhat)\right]\,,
    \label{eq:I_wtd_bound}
\end{align}
where the upper bound assumes its minimum value if the scaling parameter $s$ equals unity.
We demonstrate this bound in \Cref{fig:wtd}\,(b) for two bin widths $\Delta t$ using the data obtained by simulations. The bound is roughly a factor of 30 larger than the empirical rate function, which is due to our ansatz that reduces the high-dimensional minimization problem to a single parameter. Unfortunately, there is no obvious way to analytically find a better bound due to the complexity of the constraint \eqref{eq:EMCcond} as discussed in \Cref{sec:LDP_EPR_EMC}.

For a bound on the rate function of $\eprWTDhat$, the procedure is similar to the one above. Employing a rescaling using the parameter $s(\eprWTDhat)\equiv\eprWTDhat/\eprWTD$ results in
\begin{align}
    I(\eprWTDhat)\leq& \sum_{IJ}\int_0^\infty\dd t\,s^2\nu_I \wtdpsi{I}{J}{st}\ln\left[\frac{s\wtdpsi{I}{J}{st}}{\wtdpsi{I}{J}{t}}\right]\\ 
    &\equiv I^*\left[s(\eprWTDhat)\right]\,.
\end{align}

The upper bound on the rate function \eqref{eq:I_wtd_bound} can be used to derive a lower bound on the variance of the empirical entropy production rate by virtue of \eqref{eq:var_from_I}. We start by calculating 
\begin{equation}
    \begin{aligned}
    \partial^2_{s}\uboundbinned\!&\left[s(\eprBinnedhat)\right]_{\vert s=1}\\
    = & \left\lbrace \sum_{I,J,l}\Delta t\nu_I \frac{\left[\Psi^{(1)}_{I\to J}(t_l)\right]^2}{\binnedwtdpsi{I}{J}{t_l}} -\frac{1}{\mean{t}}\right\rbrace\,,
    \label{eq:diff_I_binned}
\end{aligned}
\end{equation}
where the expression with superscript $(1)$ is defined as
\begin{equation}
    \Psi^{(1)}_{I\to J}(st_l)\equiv\frac{1}{\Delta t}\int_{t_l-\Delta t/2}^{t_l+\Delta t/2}\dd{t} t\left[\partial_{u}\wtdpsi{I}{J}{u}\right]_{\vert u=st}\,.
    \label{eq:Psione_rescaled}
\end{equation}
Here, the order of operations is important---we scale, take the derivative and discretize in that order. Definition \eqref{eq:Psione_rescaled} is irrelevant for continuous functions, where we can directly use the derivatives of the distributions and arrive at
\begin{equation}
    \begin{aligned}
    \partial^2_{s}I^*&\left[s(\eprWTDhat)\right]_{\vert s=1}  \\
    =&\left\lbrace \sum_{IJ}\nu_I \int_0^\infty \frac{\left[t(\partial_t\wtdpsi{I}{J}{t})\right]^2}{\wtdpsi{I}{J}{t}}\dd{t} -\frac{1}{\mean{t}}\right\rbrace\,.
\end{aligned}
 \label{eq:secndD_for_var_bound}
\end{equation}
After inserting \Cref{eq:diff_I_binned,eq:secndD_for_var_bound} into \eqref{eq:var_from_I}, the lower bound for the variance of the entropy estimator reads
\begin{equation}
    \text{var}\left(\eprBinnedhat\right) \geq \frac{\left(\eprBinned\right)^2}{T\partial^2_{s}I^{\Delta t,*}(s)_{\vert s=1} }\,
    \label{eq:var_bound}
\end{equation}
and
\begin{equation}
    \text{var}\left(\eprWTDhat\right) \geq \frac{\left(\eprWTD\right)^2}{T\partial^2_{s}I^{*}(s)_{\vert s=1} }\,
    \label{eq:var_bound_cont}
\end{equation}
for the binned and the continuous case, respectively.

The bounds \eqref{eq:var_bound} and \eqref{eq:var_bound_cont} are a further main result of this paper. The intractable dependence of $\vec{\pi}$ on $\left\{\psi_{IJ}(t)\right\}$ necessitates estimating the variance of entropy production in this way, rather than directly calculating it from the moments of $\left\{\psi_{IJ}(t)\right\}$. Although the quality of our bounds may naturally vary between different systems, we emphasize their predictive nature: Given the waiting-time distributions of any system that can be described as a partially accessible Markov network, the bounds \eqref{eq:var_bound} and \eqref{eq:var_bound_cont} yield a lower bound $\sim 1/T$ on the variance of observed entropy production after a finite measurement time $T$.

\section{Conclusion and Outlook}
\label{sec:conclusion}
In this work, we have provided an accessible derivation of the rate function of both the tuple frequency in discrete-time Markov chains and the waiting-time distributions in continuous-time semi-Markov processes, which complement extant rigorous derivations, with the latter derivation applicable to systems without DTI. We have contracted these rate functions to obtain the rate functions of the entropy production in EMCs and semi-Markov processes. In the case of the EMC, we have shown how the contraction leads to transcendental equations which necessitate the numerical minimization of the rate function. For semi-Markov processes we were able to find an appropriately rescaled dynamics, which let us derive an analytical upper bound on the rate function of the entropy production. We could then use this bound to obtain a lower bound on the expected variance of the entropy production after a finite measurement time.

Future research could focus on further quantifying the quality of our analytical bounds and on finding tighter ones. Such an improvement might be achieved by introducing binning with variable bin sizes. Additionally, deriving the rate function for semi-Markov processes arising from an underlying periodically driven Markov networks remains as an open problem. When using our pedestrian's approach, infinitely many new states must be introduced for each semi-Markov state during one period of the driving. Whereas this seems to be a simple task, we expect a rigorous derivation to be less straightforward as it must deal with two time arguments and uncountable infinitely many states. Future work could also explore the impact of uncertainty introduced by other processes like an incomplete or error-prone measurement. The finite-time aspect of measurements provides another possible direction of future research. Reaching the limit in which the large-deviation theory is applicable requires a long measurement time. It would thus be interesting to study the fluctuation of waiting-time distributions and the entropy production beyond this asymptotic limit.

\section*{Acknowledgements}
We thank Jann van der Meer for insightful discussions.
%========================================[ Block separating text from the appendices ]========================================%
%
%
%
%
%
%
%
%
%
%
%
%
%
%
%
%
%
%
%
%========================================[ Block separating text from the appendices ]========================================%
\appendix\crefalias{section}{appendix}\crefalias{subsection}{appendix}

\section{Rate function for waiting-time distributions} \label{sec:proof}
The rate function for the empirical measure and the empirical flow of a semi-Markov process with DTI and a finite or a countably infinite number of states, in which waiting-time distributions may even be heavy-tailed, is known in the mathematical literature \cite{mari16,jia22}. We use the rigorously derived rate function of \citeauthor{jia22} \cite{jia22} to derive here the rate function for continuous waiting-time distributions of systems without DTI. Moreover, we comment on the general case, in which measures are needed. We thereby rigorously prove result \eqref{eq:I_wtdneu} of our pedestrian's approach in the main text and provide a sketch of the derivation for the general case, which implies result \eqref{eq:I_binnedpsi}.

Consider a semi-Markov network with a finite or countably infinite set of discrete states $V = \{x,y,\dots\}$ and DTI, which means that the transition probabilities $p_{xy}$ and the sojourn-time distribution $\psi_x(t)$ of states $x$ are independent. All sojourn-time distributions are normalized and assumed to be continuous in time, i.e., $\int_0^\infty\psi_x(t)\dd{t}=1$. Additionally, we denote their empirical version as $\hat{\psi}_x=(\vec{\hat{\psi}})_x$, the empirical transition probabilities as $q_{xy}=(\vec{q})_{xy}$ and the empirical average frequency of a state $x$ as $\hat{\nu}_x$. Assuming the sojourn-time distributions vanish for time $t=\infty$ enables us to express the rate function derived in \cite{jia22} in the somewhat simplified form
\begin{align}
I(\vec{q},\vec{\hat{\psi}}) = \sum_{x} \hat{\nu}_x\left[\sum_y q_{xy}\ln\frac{q_{xy}}{p_{xy}} +\int_0^\infty\hat{\psi}_x(t)\ln\frac{\hat{\psi}_x(t)}{\psi_x(t)}\dd{t}\right].
\label{appeq:I_jia_qpsi}
\end{align}

The rate function \eqref{appeq:I_jia_qpsi} cannot directly be used for semi-Markov processes without DTI, since it has been derived for processes that have DTI. However, we can map a semi-Markov process without DTI, which has states H,I,J,\dots, to a semi-Markov process with DTI as defined above by treating a transition $IJ$ as a state, e.g., $x=(IJ)$, and the corresponding waiting time as sojourn time of this state, as mentioned in Ref. \cite{jia22}. Importantly, sojourn-time distributions obtained this way like $\psi_x(t)$ are part of a flow into the final state of the transition $x$ within the semi-Markov network that does not require DTI. Hence, we use the identity $\nu_x = \sum_y\nu_yp_{yx}$ and reorder the double sum in the rate function \eqref{appeq:I_jia_qpsi} so that all terms describe flows into $x$. The rate function becomes
\begin{align}
I(\vec{q},\vec{\hat{\psi}}) = \sum_{y} \hat{\nu}_y\sum_x q_{yx}\left[\ln\frac{q_{yx}}{p_{yx}} +\int_0^\infty\hat{\psi}_x(t)\ln\frac{\hat{\psi}_x(t)}{\psi_x(t)}\dd{t}\right].
\label{appeq:I_jia_qpsi_inflows}
\end{align}
Using this mapping and the fact that we are dealing with inflows rather than outflows, we define $y=HI$ and $x=IJ$. These definitions imply $p_{yx}=p_{x\vert y} = p_{J\vert HI} = p_{J\vert I} = p_{IJ}$ for the transition probabilities and $\nu_y = \nu_Hp_{HI}$ for the average rate of states. Further, we have $\psi_x = \psi_{I\to J}/p_{IJ}$ and use the normalization of sojourn-time distributions $\psi_x(t)$ to combine the terms in brackets in \eqref{appeq:I_jia_qpsi_inflows}. The sums over $x$ and $y$ thereby transform to a sum over $HIJ$. Evaluating the sum over $H$ finally leads to the familiar form
\begin{equation}
    I\left[\left\{\empwtdpsi{I}{J}{t}\right\}\right] = \int_0^\infty \dd{t} \sum_{IJ} \hat{\nu}_I \empwtdpsi{I}{J}{t}\ln\frac{\empwtdpsi{I}{J}{t}}{\wtdpsi{I}{J}{t}},
    \label{eq:I_psi}
\end{equation}
which proves the result \eqref{eq:I_wtdneu} of our pedestrian's approach in the main text.

Additionally, we want to make the following remarks. Although we assume continuous sojourn- and waiting-time distributions, our transformations work in the more general case with measures too. In that case, both results \eqref{eq:I_binnedpsi} and \eqref{eq:I_wtdneu} follow directly from the transformed rate function based on measures by inserting discretized and continuous waiting-time distributions, respectively, and by adjusting the integral accordingly. Moreover, a rate function for binned waiting-time distributions is a lower bound of the rate function for continuous waiting-time distributions by virtue of Jensen's inequality. Similarly, rate functions for finely binned waiting-time distributions are always bound from below by one obtained for a coarser binning, as shown by \citeauthor{sano57} \cite{sano57}.

\bibliography{references.bib}

\end{document}